\documentclass[runningheads]{llncs}
\usepackage{graphicx}
\usepackage{amsmath,amssymb}

\DeclareMathOperator{\EX}{\mathbb{E}}
\usepackage{color}

\newcommand{\repeatthanks}{\textsuperscript{\thefootnote}}

\begin{document}
\title{Mixing-AdaSIN: Constructing a De-biased Dataset using Adaptive Structural Instance Normalization and Texture Mixing}
\titlerunning{Mixing-AdaSIN: Constructing a De-biased Dataset}
\author{
Myeongkyun Kang\inst{1} \and
Philip Chikontwe\inst{1} \and
Miguel Luna\inst{1} \and
Kyung Soo Hong\inst{2} \and
June Hong Ahn\inst{2}\thanks{Corresponding author.} \and
Sang Hyun Park\inst{1}\repeatthanks
}
\authorrunning{M. Kang et al.}
\institute{
Robotics Engineering, Daegu Gyeongbuk Institute of Science and Technology (DGIST), Daegu, Korea \\ \email{\{mkkang,shpark13135\}@dgist.ac.kr} \and
Division of Pulmonology and Allergy, Department of Internal Medicine, Regional Center for Respiratory Diseases, Yeungnam University Medical Center, College of Medicine, Yeungnam University, Daegu, Korea\\
\email{fireajh@yu.ac.kr}}
\maketitle              % typeset the header of the contribution
\begin{abstract}
Following the pandemic outbreak, several works have proposed to diagnose COVID-19 with deep learning in computed tomography (CT); reporting performance on-par with experts. However, models trained/tested on the same in-distribution data may rely on the inherent data biases for successful prediction, failing to generalize on out-of-distribution samples or CT with different scanning protocols. Early attempts have partly addressed bias-mitigation and generalization through augmentation or re-sampling, but are still limited by collection costs and the difficulty of quantifying bias in medical images. In this work, we propose Mixing-AdaSIN; a bias mitigation method that uses a generative model to generate de-biased images by \textit{mixing} texture information between different labeled CT scans with semantically similar features. Here, we use Adaptive Structural Instance Normalization (AdaSIN) to enhance de-biasing generation quality and guarantee structural consistency. Following, a classifier trained with the generated images learns to correctly predict the label without bias and generalizes better. To demonstrate the efficacy of our method, we construct a biased COVID-19 vs. bacterial pneumonia dataset based on CT protocols and compare with existing state-of-the-art de-biasing methods. Our experiments show that classifiers trained with de-biased generated images report improved in-distribution performance and generalization on an external COVID-19 dataset.

%\keywords{Bias \and Debiasing \and COVID-19 \and Bacterial pneumonia \and Image Translation \and Image Generation.}

\end{abstract}

\section{Introduction}
Recently, several methods have been proposed to improve COVID-19 patient diagnosis or treatment planning using CT scans~\cite{kang2021quantitative,li2020using,wang2020covid,wang2020contrastive}. The most methods are often trained and evaluated on single source data; producing models that exploit underlying biases in the data for better predictions. Yet, they fail to generalize when the bias shifts in external data reporting lower performance. This has critical implications, especially in medical imaging where biases are hard to define or accurately quantify. To address this, extensive data augmentation~\cite{geirhos2018imagenet} or re-sampling is often employed; though is still limited by collection (multi-institute data) and how to express the bias when it is unknown. Thus, there is a need for methods that mitigate bias in part or fully towards improved model performance and better generalization.  

In general, models trained on biased data achieve high accuracy and despite their capacity lack the motivation to learn the complexity of the intended task. For instance, a model trained on our in-house biased dataset with COVID-19 and bacterial pneumonia reported 97.18\% (f1-score) on the validation set, yet degrades to 33.55\% when evaluated on an unbiased test-set. Here, we believe bias may originate from varied CT protocols based on exam purpose, scanners, and contrast delivery requirements~\cite{caschera2016contrast}. Though contrast CT is a standard protocol, it is challenging for practitioners to meet requirements during the pandemic due to extra processes such as contrast agent injection and disinfection~\cite{pontone2020role,kalra2020chest}. Further, protocols may also vary for other pneumonia that exhibit similar imaging characteristics with COVID-19 in CT. Consequently, we believe biased datasets are often constructed unexpectedly and sometimes unavoidably due to the aforementioned factors.

Among the existing techniques proposed to remove model dependence on bias, augmentation is a de-facto technique for medical images; with other methods pre-defining the bias the trained model should be independent of. This assumes bias is easily defined, but one has to take extra care in the medical setup where such assumptions do not hold. To address this, we propose to construct a de-biased dataset where spurious features based on texture information become uninformative for accurate prediction. A key motivation is that accurate prediction of COVID-19 from other pneumonia's is dependent on the CT protocols related to texture features and contrast. Thus, we propose to generate COVID-19 CTs with bacterial pneumonia protocol characteristics and vice versa for bacterial pneumonia with COVID-19, respectively.     

Specifically, we propose Mixing-AdaSIN; a generative model based bias removal framework that leverages Adaptive Structural Instance Normalization (AdaSIN) and texture \textit{mixing} to generate de-biased images used to train classifiers robust to the original bias. For image generation, we employ two main components: (a) texture \textit{mixing}, which enables realistic image generation, and (b) AdaSIN, which guarantees structural consistency and prevents bias retainment in the input image via modifying the distribution of the structure feature. To prevent incorrect image generation, we first pre-train a contrastive encoder~\cite{he2020momentum} to learn key CT features and later use it to search similar image pairs for the texture \textit{mixing} step in the proposed generative framework. For evaluation, we construct biased train/validation sets based on the CT protocol and an unbiased test set from the non-overlapping CT protocols of the train/validation sets, respectively. The proposed method reports high bias mitigation performance (66.97\% to 80.97\%) and shows improved generalization performance when verified on an external dataset. The main contributions of this work are summarized as follows:

\begin{itemize}
  \item We propose a generative model that can sufficiently mitigate the bias present in the training dataset using AdaSIN and texture \textit{mixing}.
  \item We show that the use of contrastive learning for texture transfer pair selection prevents incorrect image generation.
  \item We constructed a biased COVID-19 vs. bacterial pneumonia dataset to verify bias mitigation performance. Our approach not only enabled improvements for standard classification models such as ResNet18 but also current state-of-the-art COVID-19 classification models.
  \item We also demonstrate the generalization performance of our classifier trained with the de-biased data on an external public dataset.
\end{itemize}

\section{Related Works}

\noindent{\textbf{CT based COVID-19 Classification}}. Several methods have been proposed to address this task since the inception of the pandemic~\cite{kang2021quantitative,li2020using,wang2020covid,wang2020contrastive}. For instance, Li et al.~\cite{li2020using} encode CT slices using a 2D CNN and aggregate slice predictions via max-pooling to obtain patient-level diagnosis. Wang et al.~\cite{wang2020covid} proposed COVID-Net, an approach that utilizes the long-range connectivity among slices to increase diagnostic performance. Later, Wang et al.~\cite{wang2020contrastive} further improve COVID-Net by employing batch normalization and contrastive learning to make the model robust to multi-center training. However, these models did not address the bias in the dataset and thus may fail to generalize on other datasets.

\noindent{\textbf{Bias Mitigation}}. To mitigate bias, Alvi et al.~\cite{alvi2018turning} employed a bias classifier and a confusion loss to regularize the extracted features to make them indistinguishable from a bias classifier. Kim et al.~\cite{kim2019learning} proposed to mitigate bias through mutual information minimization and the use of a gradient reversal layer. Though these methods can mitigate distinct biases such as color, they fail to mitigate bias in the medical domain since the bias from CT protocols is subtle and hard to distinguish even for humans~\cite{bissoto2020debiasing}. 

Another line of work is the augmentation based models that utilize techniques such as arbitrary style transfer. Geirhos et al.~\cite{geirhos2018imagenet} proposed shape-ResNet, an approach that finetunes a model pre-trained on AdaIN~\cite{huang2017arbitrary} generated images. Li et al.~\cite{li2020shape} proposed a multitask learning approach that enables accurate prediction by either using shape, texture, or both types of features. Though successful, a key drawback is the heavy reliance on artistic image generation techniques that may be detrimental for medical images and subsequent diagnoses. To address this, our approach is able to capture subtle differences in the image to generate consistent texture updated images. Thus, classifiers trained on the generated images can avoid subtle bias information.

\noindent{\textbf{Generative Texture Transfer}}. Methods \cite{huang2017arbitrary} and \cite{ghiasi2017exploring} both proposed to generate texture updated images based on arbitrary style transfer, with adaptive and conditional instance normalization employed to transfer texture information. CycleGAN~\cite{zhu2017unpaired} is another popular method for texture transfer that uses the idea of consistency across several model outputs. However, these techniques not only change the texture but also induce structural distortion to the outputs which may introduce new forms of bias. Recently, Park et al.~\cite{park2020swapping} achieved better results by using an autoencoder with texture swapping as well as a co-occurrence patch discriminator to capture high-level detail. In this method, the discriminator model may often change the original structural characteristics which is undesirable for medical images. Since our main objective is to maintain the structural information, we avoid techniques such as cycle consistency and patch discriminators that often produce structurally distorted images.

\begin{figure}[t]
\includegraphics[width=\textwidth]{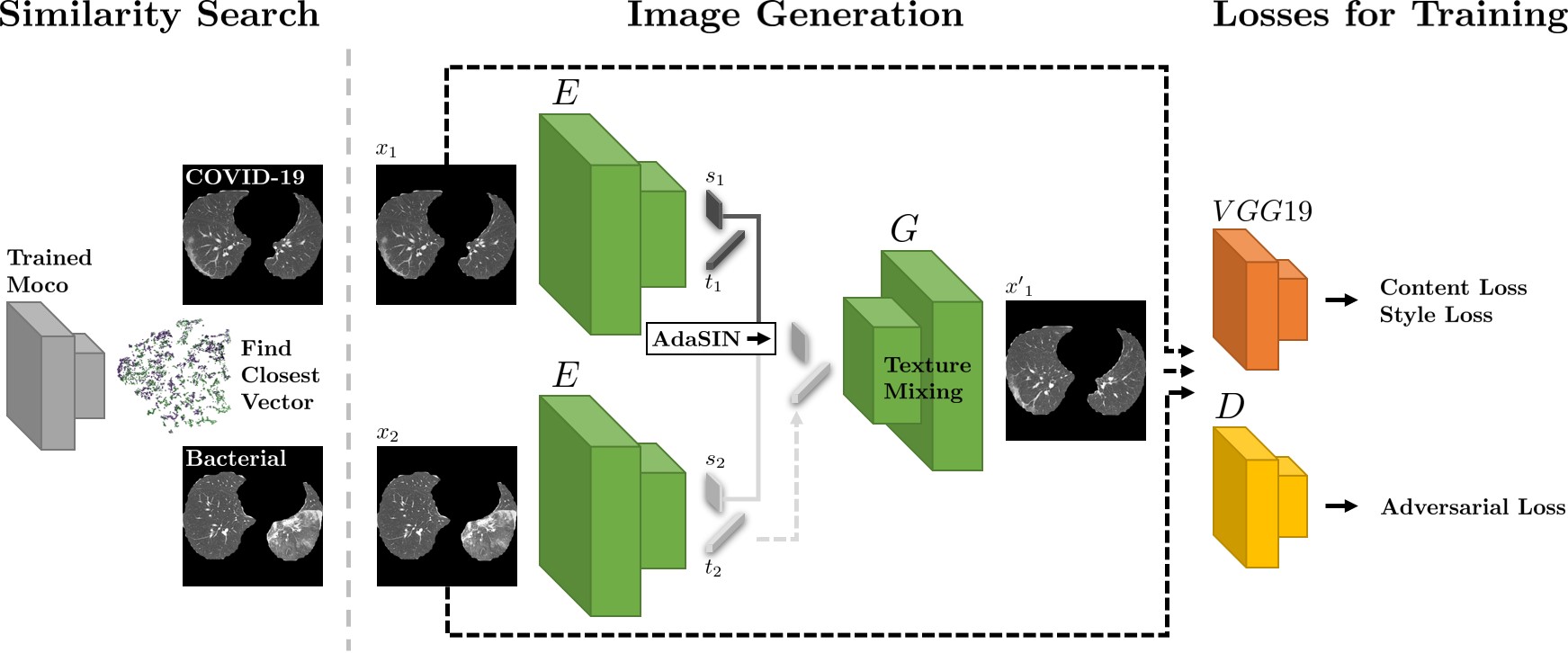}
\caption{Diagram of the proposed method.}
\label{fig1}
\end{figure}

\section{Methods}
Given a chest CT dataset $\mathcal{D}=\{X^1,...,X^n\}$ where $X^i$ is a set of 2D CT slices $X^i=\{x_1^i,...,x_m^i\}$ each with its label $y^i \in\{0,1\}$ denoting bacterial pneumonia and COVID-19 samples, respectively. Our goal is to generate a dataset $\mathcal{D}^{\prime}=\{X^{1\prime},...,X^{n\prime}\}$ where each $X^{i\prime}$ is a set of texture updated CT slices that may contain the bias information of the other label CT protocol. To achieve this, we first pre-train a contrastive encoder using $\mathcal{D}$ to learn representative slice features and then use it for slice similarity search i.e. $x_1^i,x_2^j \sim \mathcal{D},y^i \neq y^j$ with semantically similar image structures. Second, to generate $x_1^{i\prime}$ we feed searched pairs ${x_1^i,x_2^j}$ to an encoder network $E$ that outputs structural and texture features used as input for a generator network $G$. Here, AdaSIN and texture \textit{mixing} is employed on the respective features for improved generation. Lastly, following standard practice for adversarial-based methods, we also employ a discriminator network $D$ to improve the quality of generation. To enforce style and content similarity, a pre-trained VGG19~\cite{simonyan2014very} is used to optimize content loss between $x_1^{\prime}$ and $x_1$, and a style loss between $x_1^{\prime}$ and $x_2$, respectively. Through this process, we can generate images to construct $\mathcal{D}^\prime$ and then combine $\mathcal{D}$ and $\mathcal{D}^\prime$ to train a classifier that will learn to ignore bias information. The overall framework is presented in Figure~\ref{fig1} with specific details of each step categorized below.

\noindent{\textbf{Slice Similarity Search}}. As opposed to arbitrarily sampling pairs for texture transfer in our generative framework, we employ similarity based sampling of image pairs with similar structural features between the classes. Here, we pre-train a momentum contrastive network (MoCo)~\cite{he2020momentum} using the training data and find the closest slice based on the $L1$ distance. This is crucial for generation since using arbitrary image pairs for texture transfer can produce artificial images without any clinical significance. Following, we construct image pairs for the entire dataset for image generation. 

\noindent{\textbf{Image Generation with AdaSIN and Texture \textit{Mixing}}}. To generate a de-biased and texture mixed image, an encoder network $E$ takes as input the sampled image pairs to produce texture and structure features that are first modified via AdaSIN before being feed to $G$. To retain texture and structure, both are required as inputs for $G$. Specifically, the structure feature is a feature with spatial dimensions pooled at an earlier stage of $E$, whereas the texture feature is an embedding vector obtained following $1\times1$ convolution in the later stages of $E$. In addition, we believe that directly employing these features for generation can still retain the inherent bias contained in the features, thus we propose to modify the structural features via AdaIN~\cite{huang2017arbitrary}. To achieve this, we use the mean($\mu(\cdot)$) and standard deviation ($\sigma(\cdot)$) of the features before passing to $G$. Formally,
\begin{equation}
AdaSIN(s_1,s_2)=\sigma(s_2)\left(\frac{s_1-\mu(s_1)}{\sigma(s_1)}\right)+\mu(s_2),\label{eq1}
\end{equation}
where $s_1$ and $s_2$ denotes the extracted structure features of the input image pairs i.e. $x_1,x_2 \sim \mathcal{D}$. Next, $G$ takes both features for an image generation and texture transfer. Texture transfer is achieved via convolution weight modulation and demodulation introduced in \cite{karras2020analyzing}. Herein, texture information is delivered to each convolutional layer in $G$ except for the last layer.

To train the entire framework, we follow the arbitrary style transfer training procedure described in \cite{huang2017arbitrary}. A VGG19 pre-trained model is used to extract features from the input pairs which is then used in the style $L_{style}$ and content $L_{content}$ losses, respectively. The style loss minimizing the mean squared error (MSE) between the generated output and the texture image, whereas the content loss minizing the MSE between generated output and the structure image~\cite{li2017demystifying}. Further, we use an adversarial loss to improve the image generation quality via $L_{GAN}(G,D)=-\EX_{x_1,x_2 \sim \mathcal{D}}[D(G(AdaSIN(s_1,s_2),t_2))]$, with regularization (non-saturation loss) omitted for simplicity~\cite{gulrajani2017improved,karras2020analyzing}. The final loss function is defined as:
\begin{equation}
L_{total}=L_{content}+ \lambda L_{style}+L_{GAN}.\label{eq2}
\end{equation}
\noindent{\textbf{{Implementation Details}}. A Mask-cascade-RCNN-ResNeSt-200~\cite{zhang2020resnest} with deformable convolution neural network (DCN)~\cite{dai2017deformable} was employed to extract the lung and lesion regions in the CT scans to mask out the non-lung regions and excluding slices that do not contain a lesion. The contrastive slice encoder was trained for 200 epochs whereas the generative model was trained with a batch size of 8 for 40,000 steps with Adam optimizer and learning rate of 0.002. Further, the discriminator $D$ follows Karras et al.'s~\cite{karras2018progressive,karras2020analyzing} implementation and we empirically set $\lambda=10$.

\begin{table}[t]
\centering
\caption{The statistics and bias information (CT protocol) of the train/validation/test dataset.}
\label{tab1}
\begin{tabular}{p{1cm}|p{1.5cm}|p{3.8cm}|p{1.5cm}|p{3.8cm}}
\hline
\bfseries Split & \bfseries COVID & \bfseries CT protocol & \bfseries Bacterial & \bfseries CT protocol \\
\hline
Train & 42 & Non-Contrast (kVp: 120). & 60 & Contrast (kVp: 100). \\
\hline
Val & 10 & Non-Contrast (kVp: 120). & 14 & Contrast (kVp: 100). \\
\hline
Test & 21 & Various CT protocols from different scanners: Contrast (18), etc (3).  & 23 & Various CT protocols from different scanners: Non-Contrast (18), etc (5). \\
\hline
\end{tabular}
\end{table}

\section{Experiments and Results}
\subsubsection{Experiments Settings} For evaluation, we constructed an in-house biased dataset for training/validation and an unbiased testing dataset. First, we extracted the CT protocol per scan using metadata in the DICOM files and create splits COVID-19 and bacterial pneumonia based on the protocol. The dataset and bias information are shown in Table~\ref{tab1}.

To evaluate the effect of bias mitigation using the generated images for classification, a pre-trained ResNet18 and recent COVID-19 classification models~\cite{he2016deep,li2020using,wang2020covid,wang2020contrastive} i.e. COVNet, COVID-Net, Contrastive-COVIDNet were compared. 
The models were trained for 100 epochs with a batch size 64 and a learning rate 0.001. We also applied random crops, horizontal flips and intensity augmentations i.e. (brightness and contrast) as a baseline augmentation technique. Performance comparison of our approach against recent state-of-the-art non-generation based bias mitigation methods~\cite{kim2019learning,alvi2018turning} applied for natural image classification is also reported. To verify the effectiveness of the proposed method, we include comparisons against a commonly used arbitrary style transfer model i.e. AdaIN, and the current state-of-the-art generation method i.e. swapping-autoencoder~\cite{huang2017arbitrary,park2020swapping}. For a fair comparison of the generation based methods, the same texture pairs were utilized for a generation. Also, training and validation were performed three times for all methods with final average performance reported. 

\subsubsection{Results on Internal Dataset} In Table~\ref{tab2}, we present the evaluation results on the biased COVID-19 vs. bacterial pneumonia dataset. Initially, the model shows high f1-score on the  validation dataset i.e. 97.18\%, yet significantly drops to 33.55\% on the unbiased test set. This shows that the classifier makes predictions based on the bias information.  
The results of the learning-based models i.e. Learning not to learn, and Blind eye show no considerable performance improvements and highlight the failure to mitigate the bias, especially for medical domain images as reported in \cite{bissoto2020debiasing}. Further, these methods were proposed to remove bias that is distinctly recognizable in the image such as color. Capturing and mitigating the subtle bias difference in the medical image is considerably harder for such techniques.

\begin{table}[t]
	\centering
	\caption{The test results (f1-score) on biased COVID-19 vs. bacterial pneumonia dataset. The results of validation dataset are written as Val. The model without intensity augmentation is denoted w/o aug.}
	\label{tab2}
	\begin{tabular}{p{2.3cm}|p{2.18cm}p{2.18cm}p{2.22cm}p{2.7cm}}
		\hline
		& \bfseries ResNet18 \cite{he2016deep} & \bfseries COVNet \cite{li2020using} & \bfseries COVID-Net \cite{wang2020covid} & \bfseries Contrastive COVID-Net \cite{wang2020contrastive} \\
		\hline
		Val w/o aug & 97.18 & 89.33 & 78.30 & 93.31 \\
		\hline
		Base & 66.97 & 63.41 & 41.89 & 53.37 \\
		Base w/o aug & 33.55 & 59.52 & 43.99 & 35.64 \\
		\hline
		LNTL~\cite{kim2019learning} & 31.87 & - & - & - \\
		Blind eye~\cite{alvi2018turning} & 32.72 & - & - & - \\
		\hline
		AdaIN~\cite{huang2017arbitrary} & 75.71 & 66.35 & 51.35 & 65.76 \\
		Swap~\cite{park2020swapping} & 76.84 & 67.64 & 56.28 & 62.60 \\
		Mixing-AdaSIN & \bfseries 80.97 & \bfseries 74.61 & \bfseries61.57 & \bfseries 66.16 \\
		\hline
	\end{tabular}
\end{table}

\begin{figure}[b]
	\includegraphics[width=\textwidth]{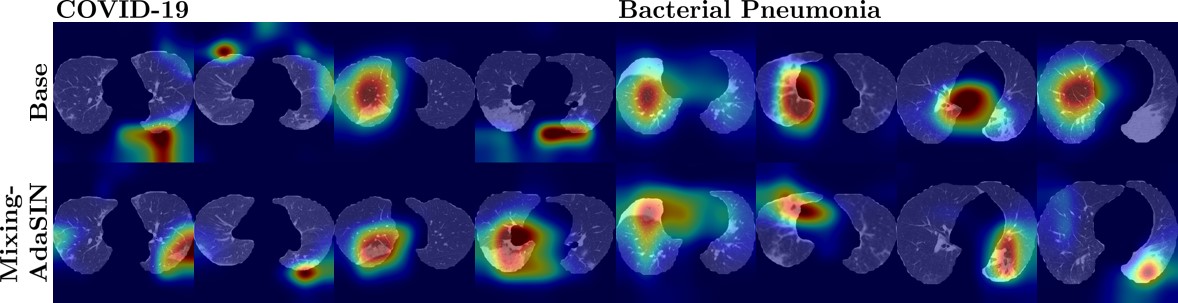}
	\caption{Grad-CAM~\cite{selvaraju2017grad} visualizations of the ResNet18~\cite{he2016deep} classifier. Herein, Grad-CAM of the base classifier pointed the normal lung area. On the other hand, the classifier trained with Mixing-AdaSIN pointed the lesion correctly. Hence, the model with debiasing can be more generalized.}
	\label{fig2}
\end{figure}

Among generation based methods, the proposed method reports the best performance. Even though AdaIN can transfer texture well, the quality of the generated image is extremely low. Consequently, this inhibits classifier training as shown by the limited performance improvements. Though swapping-autoencoder updates the texture with high quality image generation results, two major drawbacks were noted: (i) the generated image still retains bias, and (ii) it distorts key representative characteristics by artificial translation of lesions i.e. changing the lesion of COVID-19 to appear as bacterial pneumonia. Such phenomena may be due to the direct usage of structure features, and use of the co-occurrence discriminator which leads to structural information deformation. On the other hand, our model employs two modules i.e. boosting a high-quality image generation via texture \textit{mixing}; and minimizing the bias artifact transfer through an AdaSIN. We consider these techniques as instrumental in mitigating bias. In addition, the classifier with a de-biased method employed a more generalized feature as shown in Figure~\ref{fig2}. The Grad-CAM~\cite{selvaraju2017grad} of the trained classifier pointed to the lesion more correctly, thus a better generalization performance is expected.

Our proposed method can be easily applied to existing CT based COVID-19 classification models. In particular, COVNet reported 74.61\% f1-score which represents successful mitigation of bias artifacts. However, COVID-Net and Contrastive COVID-Net showed relatively low accuracy, mainly due to slight differences in training details and architectures. Also, due to the long-range connectivity in the models, reliance on bias information is heavily induced during training.

\subsubsection{Results on External Dataset} To verify the generalization efficacy of our trained classifier on external data, we employ the publicly available MosMed dataset~\cite{morozov2020mosmeddata}. It consists of 1110 CT scans from COVID-19 positive patients. However, the dataset contains CT scans that are not consistent with pneumonia observed in our original dataset. Thus, we selected scans of severe and critical patients only to evaluate the trained models. In addition, since we trained three classifiers from an internal experiment, we tested each classifier three times and final average performance reported. In Table \ref{tab3}, results are fairly consistent with improvements shown on the internal dataset evaluation. Our model shows a significant improvement over the baseline with +1\% gain over swapping-autoencoder. More importantly, even though the classifier has not observed the CT samples with a different protocol, performance was still consistent verifying the utility of the proposed de-biasing technique.

\begin{table}[t]
	\centering
	\caption{The test results (accuracy) of the ResNet18~\cite{he2016deep} on the external COVID-19 dataset.}
	\label{tab3}
	\begin{tabular}{p{1.4cm}|p{1.7cm}p{2.6cm}|p{1.8cm}p{1.8cm}p{2.3cm}}
		\hline
		& Base & Base w/o aug & AdaIN~\cite{huang2017arbitrary} & Swap~\cite{park2020swapping} & Mixing-AdaSIN \\
		\hline
		Accuracy & 24.11 & 34.04 & 50.35 & 76.60 & \bfseries 77.30 \\
		\hline
	\end{tabular}
\end{table}

\section{Conclusion}
In this work, we have proposed a novel methodology to train a COVID-19 vs. bacterial pneumonia classifier that is robust to bias information present on training data. We constructed an in-house biased training dataset in conjunction with an unbiased testing dataset and proved that our method allowed the classifier to learn the appropriate features to correctly predict the labels without considering bias information and achieved better generalization. All of this was possible thanks to an adequate image generation design that relies on two major components: (a) texture \textit{mixing}, which enables realistic image generation, and (b) AdaSIN, which prevents bias flow from the input to the output image in the generation stage, while maintaining structural consistency. We proved the benefits of our pipeline by achieving the best bias mitigation performance when compared to other related methods in both our in-house dataset as well as in an external dataset. Considering that biases can be easily included when constructing datasets, we hope that our findings help to improve performance in various medical tasks.

\subsubsection{Acknowledgment.}
This research was funded by the National Research Foundation of Korea (NRF) grant funded by the Korean Government (MSIT) (No.201 9R1C1C1008727).

\bibliographystyle{splncs04}
\bibliography{ref}

\end{document}